\documentclass[12 pt]{article}
\usepackage[utf8]{inputenc}
\usepackage{authblk}
\usepackage{graphicx}
\usepackage{caption}
\usepackage{subcaption}
\usepackage[margin=0.8in]{geometry}
\usepackage[usestackEOL]{stackengine}
\usepackage[%
    colorlinks=true,
    linkcolor=red
]{hyperref}
\usepackage{setspace}
\usepackage[section]{placeins}
\title{Electron spin mediated distortion in metallic systems}
\author[1,2,3,*]{G. Anand}
\author[4]{Markus Eisenbach}
\author[2]{Russell Goodall}
\author[2]{Colin L. Freeman}
\affil[1]{{Department of Metallurgy and Materials Engineering, Indian Institute of Engineering Science and Technology-Shibpur, Howrah, WB, India}}
\affil[2]{Department of Materials Science and Engineering, University of Sheffield, UK}
\affil[3]{Warwick Centre for Predictive Modelling, University of Warwick, UK}
\affil[4]{Scientific Computing Group, Centre for Computational Sciences, Oak Ridge National Laboratory, Oak Ridge, TN, USA}
\affil[*]{ganand@metal.iiests.ac.in}
\date{}
\begin{document}
\maketitle

\begin{figure}[htb!]
    \centering
    \includegraphics[width=\textwidth]{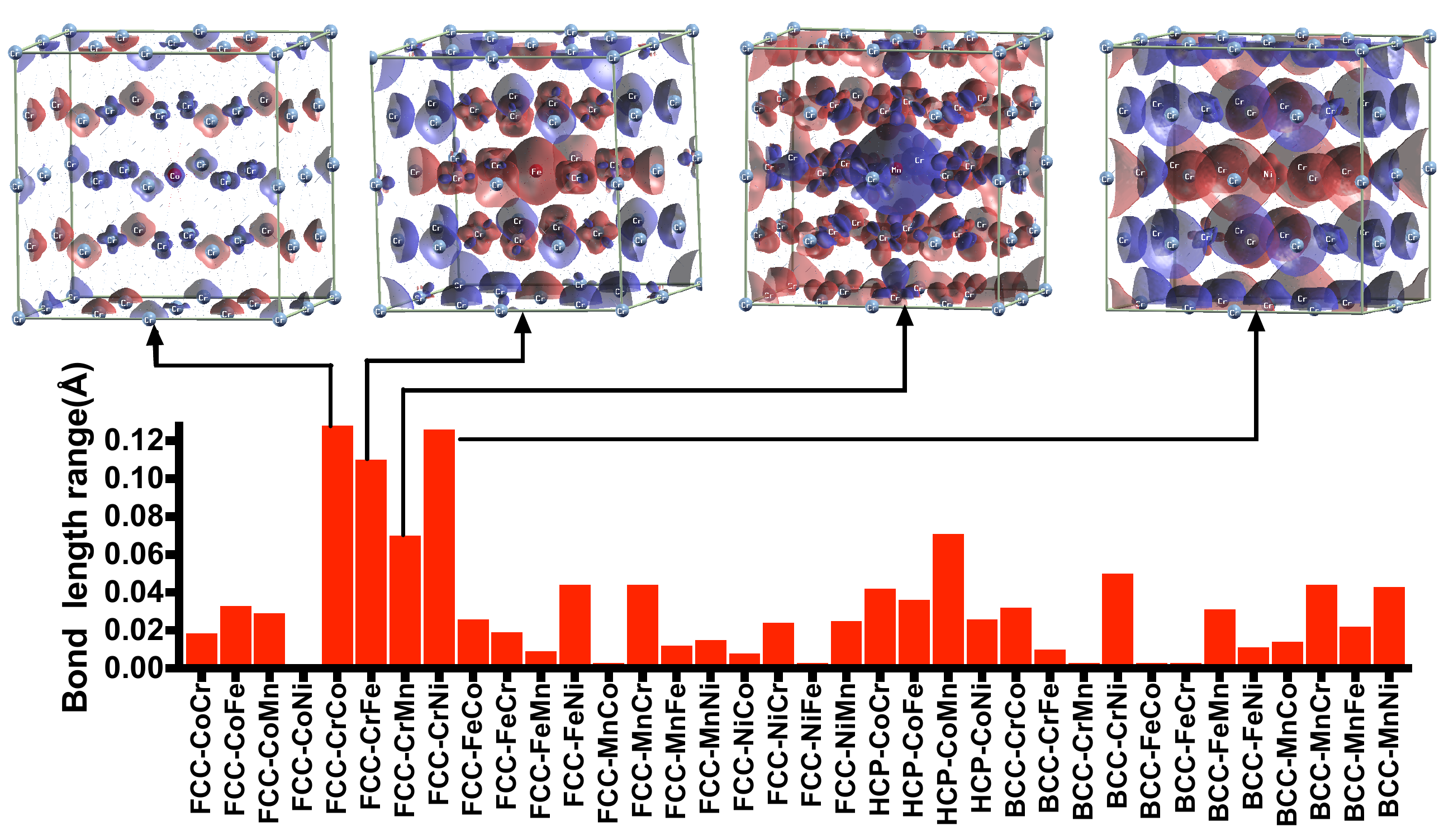}
    \caption{Variation in the bond distortion for various cases, spin iso-surface for Cr matrix has been added showing Cr with both up and down spin.
    }
    \label{fig1:distortion}
\end{figure}

\begin{figure}
    \centering
    \begin{subfigure}[b]{0.3\textwidth}
        \includegraphics[width=\textwidth]{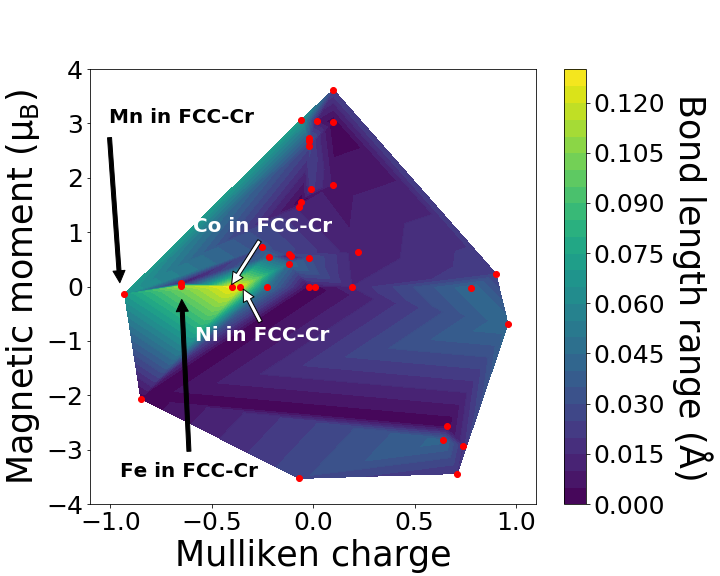}
        \caption{}
        \label{fig2:charge_mm_distortion}
    \end{subfigure}%
    \begin{subfigure}[b]{0.7\textwidth}
        \includegraphics[width=\textwidth]{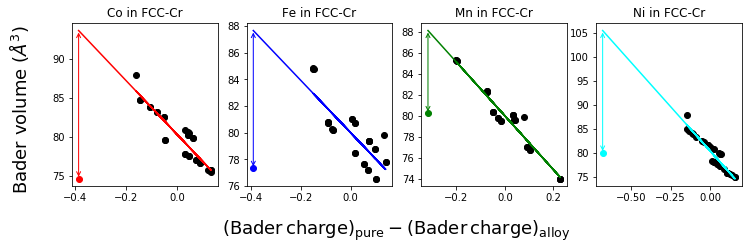}
        \caption{}
        \label{fig2:bader_Cr}
    \end{subfigure}
    
    \begin{subfigure}[b]{0.3\textwidth}
        \includegraphics[width=\textwidth]{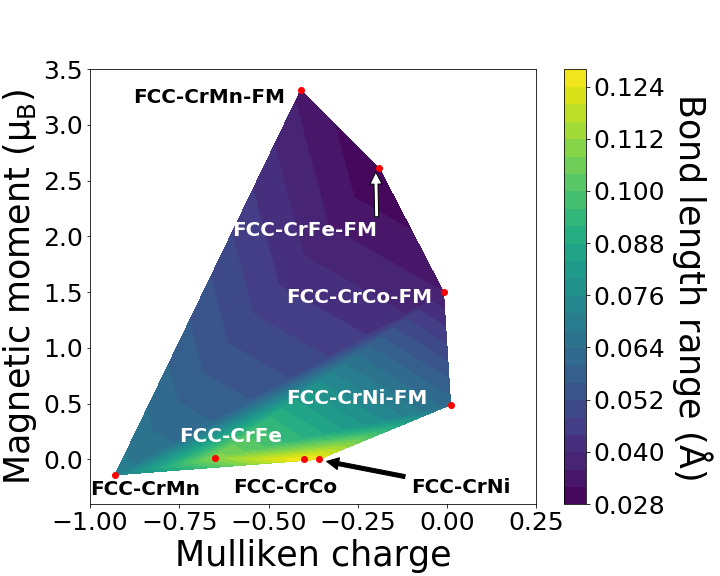}
        \caption{}
        \label{fig2:charge_mm_fm}
    \end{subfigure}%
    \begin{subfigure}[b]{0.7\textwidth}
        \includegraphics[width=\textwidth]{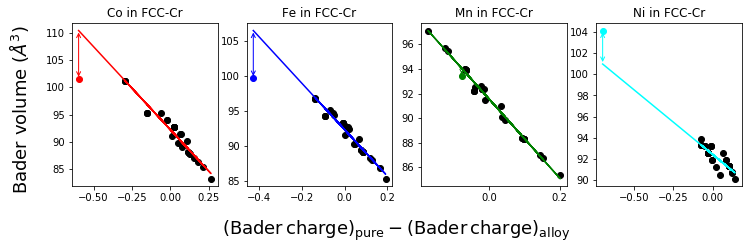} 
        \caption{}
        \label{fig2:bader_Cr_fm}
    \end{subfigure}
    \begin{subfigure}[b]{0.5\textwidth}
        \centering
        \includegraphics[width=0.7\textwidth]{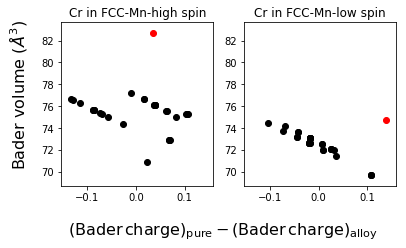}
        \caption{}
        \label{fig2:Mn-vol-change}
    \end{subfigure}
    \caption{
    (\subref{fig2:charge_mm_distortion}) Variation of Mulliken charge, magnetic moment of impurity and distortion in the supercell
    (\subref{fig2:bader_Cr}) Bader volume associated with net Bader charge of impurity and matrix
    (\subref{fig2:charge_mm_fm}) Change in Mulliken charge, magnetic moment of the impurity and consequent change in distortion for ferromagnetic system
    (\subref{fig2:bader_Cr_fm}) Bader volume associated with net Bader charge for impurity and matrix elements for ferromagnetic case
    (\subref{fig2:Mn-vol-change}) Bader volume associated with Mn and impurity Cr atoms with respect to net charge for high spin (high magnetic moment) and low spin (low magnetic moment) case.
    }
\end{figure}

\begin{figure}
    \centering
    \includegraphics[width=0.8\textwidth]{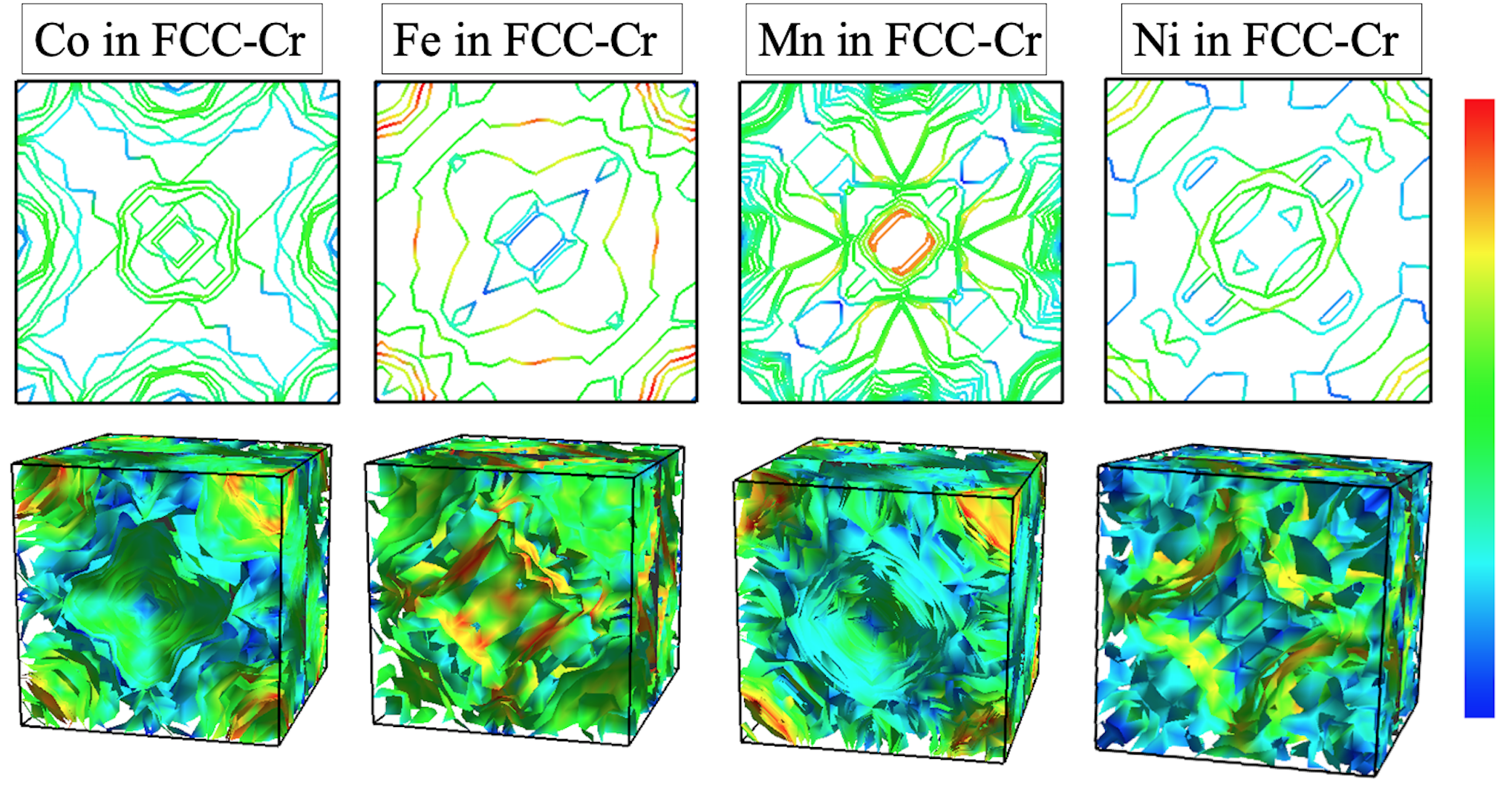}
    \stackinset{c}{+1.55in}{t}{+0.70in}{\includegraphics[width=2.in]{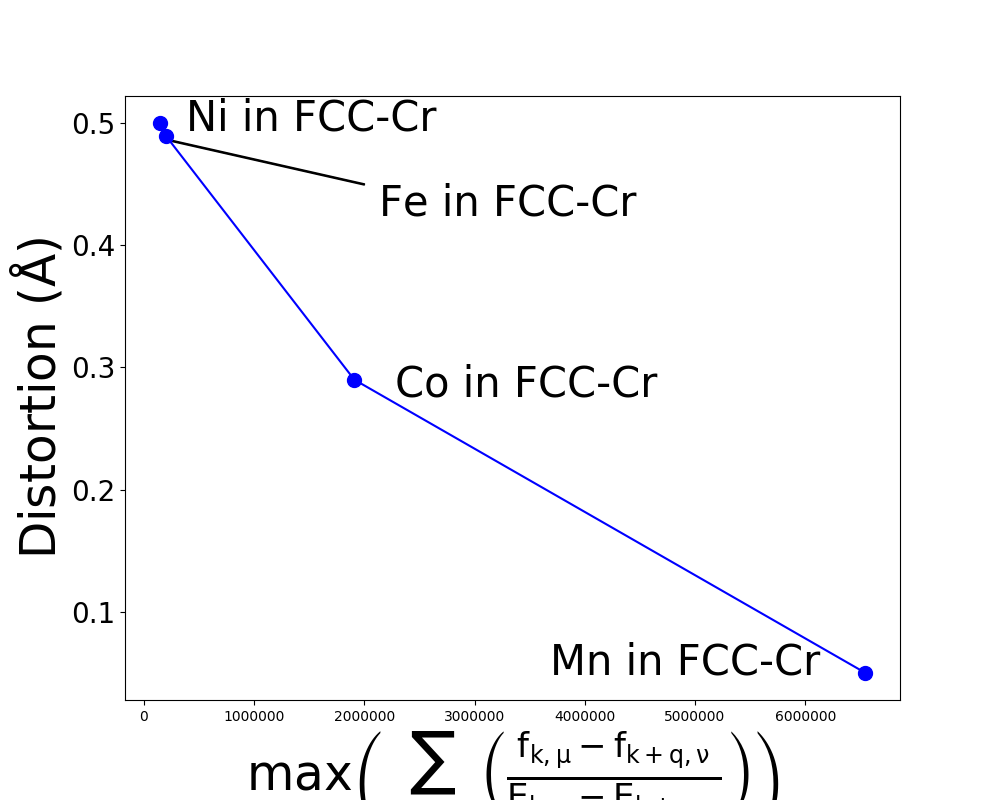}}{%
    \includegraphics[width=\textwidth]{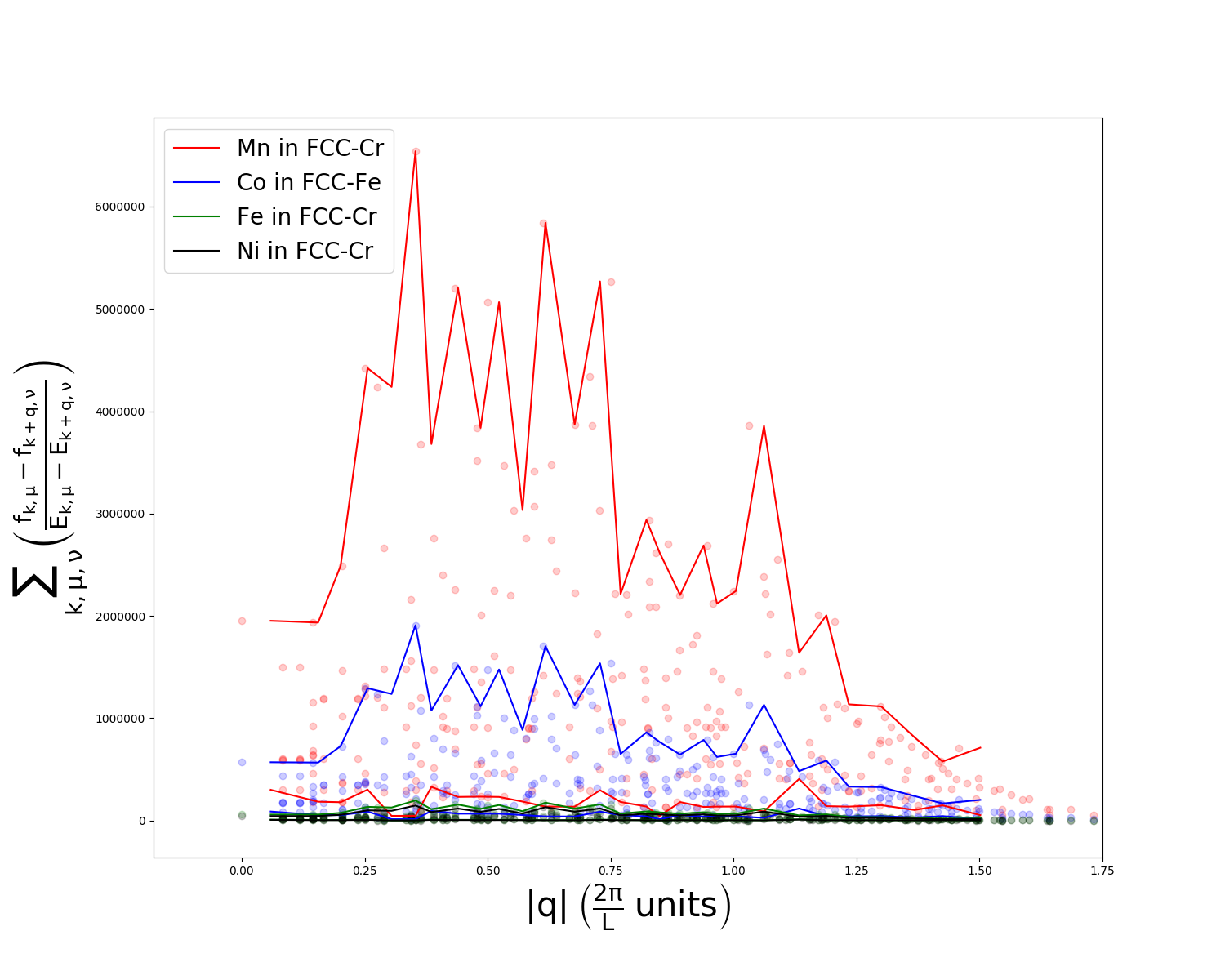}}
    \caption{Fermi surface topology for Co, Fe, Mn and Ni in FCC-Cr, along with response function for each of these cases. The inset in the figure show the maximum value of response function with distortion in the supercell.}
    \label{fig3:respose_all}
\end{figure}

\doublespacing
\textbf{The deviation of positions of atoms from their ideal lattice sites in crystalline solid state systems causes distortion and can  lead  to  variation  in  structural \cite{Song2017} and functional properties \cite{Berardan2017}. Distortion in molecular systems has been traditionally understood in the term of Jahn-Teller distortion \cite{Independent2014}, while for the one-dimensional chain of metals, the Peierls distortion mechanism has been proposed \cite{Hoffmann1987}. In real three dimensional metallic systems, a fundamental description of the distortion is missing, which we need to design alloys with attractive structural properties. The present investigation presents the evolution of distortion in metallic systems in terms of magnetovolume effects arising due to magnetic ground-state of the system. Particularly the significant distortion in the case of Cr due to presence of other transition metals is seen. Cr with transition metals as a candidate system to study the correlation between charge disproportion, spin-flip, magnetovolume effects and Fermi surface nesting. This study provides a novel explanation of strengthening effect of Cr observed in alloys due to its unique magnetic properties.} \\
Distortion in metallic systems is a topic of debate in the context of multi-component concentrated alloys or so called "high-entropy alloys (HEA)" \cite{George2013, Lee2018, Choi2018, Tong2018, Song2017a, Owen2018}. Such distortions are known to influence the structural properties of these alloys \cite{He2018, Feng2018}. Local lattice distortion is instrumental in the determination of elastic properties, where size misfit plays the crucial role by modifying the elastic properties of metallic alloys\cite{Kim2019}. Accurate determination of such properties is a key part in developing predictive models for designing newer alloys \cite{Leyson2010}. In view of delineating the elemental influence, we have studied the evolution of distortion with the impurity-in-matrix approach \cite{Morinaga2019} with elements forming the CoCrFeMnNi HEA. We generated an FCC crystal structure for Co, Cr, Fe, Mn and Ni, while a BCC structure was used for Cr, Fe, and Mn. A HCP matrix was generated for Co only. Other elements were added as an impurity in each of the above-stated matrices in the centre of the supercell. Geometry optimisation in Density Functional Theory (DFT) was carried out to determine the equilibrium bond-length between various atoms in the supercell. Fig. \ref{fig1:distortion} shows the histogram of bond-length range for various cases. The bond-length range was determined as the difference between maximum and minimum of the equilibrium bond-length values present in the supercell. We considered this bond-length range as a measure of the distortion in the supercell. In each case it was bond-length values for each atom with its first nearest-neighbour which were determined. It is clear from our results that FCC-Cr with impurities tends to have greater distortion in comparison with the other cases examined. It should be noted that Cr and Mn in multi-component alloys are known to cause maximum distortion \cite{Oh2016}. We have plotted spin-isosurfaces for these cases and we found finite spin-polarisation in these systems. It can be seen that Cr at  the first nearest-neighbour position tends to have both up and down spin associated with it. Electronic redistribution, spin-flip and change in the bonding structure are correlated in transition metal based systems \cite{Wang2015}.\\
We have additionally looked into the charge disproportion and magnetic moment of the impurity atom in the matrix and plotted it with respect to the distortion in the supercell. It can be seen in Fig. \ref{fig2:charge_mm_distortion} that in the case of FCC-Cr, the impurity atom gains electronic charge and the magnetic moment associated with impurities remains close to zero. Such charge-transfer has been reported to be important in atomic-level stress \cite{Oh2019, Tong2019}. In view of quantifying the influence of the charge disproportion, we have carried out Bader charge analysis, where the net charge on the atom was calculated using Henkelman's grid-based algorithm \cite{Tang2009}. After calculation of Bader charge on all atoms in the supercell, we subtracted the Bader charge on a particular atom from the original electronic charge in the pseudopotential to determine the net gain or loss of charge on each atom. It can be seen that impurity atoms in FCC-Cr, tend to gain electronic charge, which is not in the agreement with Pauling's electronegativity scale for Mn in FCC-Cr but rather, Allen's scale provides the correct electronegativity trend for all the cases (see Table 1 in the supplementary information for electronegativity values). The Allen's scale of electronegativity has been found to be successful for HEA \cite{Poletti2014}. Even with net electronic gain on the impurity atoms, interestingly the Bader volume occupied by the impurity atoms seems to be shrinking (Fig. \ref{fig2:bader_Cr}) and such volume misfit has been correlated with distortion \cite{Nohring2019}. Recent experimental work has also shown that Cr influences the lattice parameter of CoCrFeNi alloy most in comparison with Co, Fe and Ni and hence it exhibits the strongest lattice distortion effect in this alloy \cite{Wang2019}. It should be further noted that the impurity-first nearest neighbour bond-length is lower than the bond-length of Cr-Cr pairs.\\
So to determine the influence of the magnetism on the evolution of distortion, we carried out constrained spin-polarised calculations for the case, where we saw significant distortion (\emph{i.e.,} FCC-Cr with impurities). In these calculations, we forced ferromagnetism on the system by constraining the supercell to have a total initial magnetic moment. We confirmed the imposition of ferromagnetism by determining total electronic density of states (DOS) (See Figure 1 in the Supplementary Information (SI)), which shows the exchange split in up and down spin for constrained ferromagnetic calculations. It can be clearly seen that forcing ferromagnetism causes the reduction of distortion (Fig. \ref{fig2:charge_mm_fm}) as well as reduction in volume-shrinkage associated with the impurity atoms in the supercell (Fig. \ref{fig2:bader_Cr_fm}). Such observation points towards a correlation between volume shrinkage and distortion in such systems. The correlation of volume shrinkage and reduction in the bond-length between substitutional solute and Cr explains the  local distortion (or negative strain) observed when 3d transition metals are alloyed with another transition metal \cite{Morinaga2019}. Also, it can be seen that forcing ferromagnetism in FCC-Cr system leads to the lower charge disproportion on the impurity atom and higher magnetic moment. Additionally, we found two magnetic ground-states for FCC-MnCr (\emph{i.e.,} Cr in FCC-Mn); first with a high magnetic moment associated with Cr and Mn (high spin state) with higher distortion $\mathrm{0.12\;\AA}$ and a second one with lower magnetic moment with lower distortion $\mathrm{(0.036\; \AA)}$. The characteristic of distortion in the case of FCC-MnCr (\emph{i.e.}, Cr impurity in FCC-Mn) is different from as that seen in the case of FCC-Cr matrix. In the case of the FCC-Cr matrix, distortion takes place at impurity-Cr bonds, while in the case of FCC-MnCr, Mn-Mn interaction causes the distortion. The Bader analysis similarly showed the discrepancy in the volume associated with Mn, which is minimised in the case of low magnetic moment case (Fig. \ref{fig2:Mn-vol-change}).\\
As stated above, substitutional alloying in FCC-Cr leads to finite spin polarisation in the system or the magnetic moment associated with the FCC-Cr matrix atoms and impurity atom has some finite value. The maximum and minimum moment associated with FCC-Cr with impurity supercell is -0.004 to 0.001 $\mathrm{\mu_B}$ for FCC-CrCo (Co impurity in FCC-Cr), -0.018 to 0.038 $\mathrm{\mu_B}$ for FCC-CrFe (Fe impurity in FCC-Cr), -0.168 to 0.016 $\mathrm{\mu_B}$ for FCC-CrMn (Mn impurity in FCC-Cr) and -0.0002 to 0.0004 $\mathrm{\mu_B}$ for FCC-CrNi (Ni impurity in FCC-Cr).    
We visualised the Kohn-Sham orbitals in the plane of the impurity. We found the fluctuation in the magnetic moment of Kohn-Sham states between the impurity and Cr in the FCC-Cr matrix. Such fluctuation is absent in the case of ferromagnetic systems (see Figure 2(a) and 3 in SI). Additionally such fluctuation was also seen in the case of high spin FCC-MnCr, where significant distortion is seen (Fig. 2 (b) and (c) in SI). To ensure that such fluctuation might not be due to numerical noise, we studied the fluctuation at various k-points (Fig. 5 in SI). It is clear that the ratio of the magnetic moment of conduction Kohn-Sham states and that of impurity atom remains close to $\mathrm{10^{-3}}$. We additionally contracted and expanded the supercell of the FCC-CrCo system and carried out electronic minimisation to the effect of lattice movement on the appearance of the fluctuation in the magnetic moment of the conduction Kohn-Sham states.  We observed that when the system was contracted by 0.93 of the equilibrium volume, the spin fluctuations were still present suggesting a link to the relaxation of the atoms towards each other in the general case.  Expansion of the system removed the presence of the fluctuation again suggesting that this effect is linked to the localised distortion of atoms moving closer together within the cell.  Therefore the fluctuation appears to be a screening behaviour between the conduction Kohn-Sham states. It should also be noted that such fluctuations are closely correlated with the atomic movement.
In view of influence of conduction Kohn-Sham states in the above-stated cases, we generated the Fermi surfaces for FCC-CrCo, FCC-CrFe, FCC-CrMn and FCC-CrNi. These calculations were carried out using a larger supercell with 108 atoms. We observed the parallel sheets of the Fermi surfaces (Fig. \ref{fig3:respose_all}) in each case. In Figure \ref{fig3:respose_all} we also show a two-dimensional cross section of the Fermi surfaces for all cases which have been obtained from sectioning the three-dimensional surface parallel to the (020) plane. . The parallel sheets to Fermi surface are quantified in terms of vector for which the sheets are parallel. This phenomena is known as Fermi surface nesting and the above-stated vector is the nesting vector ($\mathrm{q}$). To quantify the Fermi surface nesting, we determined the response function  ($\mathrm{\chi(q)}$) \cite{Gupta1971} as,
\begin{equation}
\mathrm{\chi(q)=\sum_{k,\mu,\nu}\left({\frac{f_{k,\mu}-f_{k+q,\nu}}{E_{k,\mu}-E_{k+q,\nu}}}\right)}
\end{equation}
where, $\mathrm{\mu}$ and  $\mathrm{\nu}$ are band indices, while $\mathrm{f}$ is occupancy factor, which is zero above the Fermi energy and one below the Fermi energy. The results across the cell are shown Fig. \ref{fig3:respose_all}. The inset of Fig. \ref{fig3:respose_all} shows the linear correlation of $\mathrm{max(\chi(q)}$ with distortion, where systems with lower Fermi surface nesting have higher distortion.  In pure Cr in the BCC state, Overhauser predicted spin density wave (SDW) \cite{Overhauser1962}, while Lomer suggested that the SDW is enhanced if Fermi surface nesting is present \cite{Lomer1962}. Also, Marcus \emph{et. al.,} showed that strain waves (sinusoidal variation of strain) due to the magneto-volume effect may also stabilise SDW \cite{Marcus1998}. So, it can be stated that Fermi surface nesting and distortion in pure Cr are positively correlated. In our investigation for FCC-Cr with impurity, we have found a negative correlation between Fermi surface nesting and distortion. It should be noted that in our case, Co, Fe and Ni in FCC-Cr lead to the ferromagnetic ordering with fluctuating spin at the interstices, while AFM ordering is present in the case of Mn in FCC-Cr. It is clear that spin fluctuations are mediating the magneto-volume effects (\emph{i.e.,} volume shrinkage). \\ 
In summary, the present investigation shows the correlation between charge disproportion, spin-flip, Fermi surface nesting with the distortion in metallic systems. We have observed the volume shrinkage of the valence electron cloud, which leads to the reduction of equilibrium bond-length between metallic elements. Such volume shrinkage is related to the magnetic ground-state along with the  screening effects of conduction electrons. It has also been shown that distortion and Fermi surface nesting are negatively correlated. This study provides a fundamental description of distortion in three dimensional metallic systems which will be relevant for the design of newer alloys with attractive structural properties.   

\bibliography{main}
\bibliographystyle{naturemag}

\section*{Method}
DFT calculations were performed using the CASTEP code \cite{Clark2005}, which uses a plane- wave expansion of one-electron wave functions. Ultrasoft psuedopotential were used to define the electron-ion interaction \cite{Vanderbilt1990}. The electron exchange-correlation was defined using the generalised gradient approximation \cite{Hamann1979} with a Perdew-Burke-Ernzerhof functional \cite{Perdew1996}. Spin-polarised calculations were carried out to introduce magnetism. Geometric optimisation was carried out in all the cases to obtain the statically relaxed structure with tolerance values for ionic displacement, force on ions, stress on ions set to 0.001 $\mathrm{\AA}$ , 0.05 $\mathrm{\AA}$, and 0.1 GPa, respectively. The geometric relaxation was carried out without imposing any constraints to determine the magnetic ground state of the system. The DFT calculations were performed with the supercell approach with the impurity-in-matrix methodology which means that a different element is added to the centre of the supercell of a pure elemental system. The FCC structure was modelled for Co, Cr, Fe, Mn and Ni, while a BCC structure for Cr, Fe and Mn was generated. The HCP structure for Mn was generated as well. A 2X2X2 supercell was generated for BCC, FCC and HCP cases, containing 16, 32 and 16 atoms respectively. In each of these cases, the remaining elements are added as a perturbation substitutionally to determine their individual effect on the bond-lengths to characterise the influence of individual elements on distortion in alloys. The plane wave cut-off ( $ \mathrm{E_{cut}} $) for Co, Cr, Fe, Mn and Ni were determined to be 500 eV, 700 eV, 500 eV, 700 eV and 500 eV, respectively. The number of k-points for Brillouin-zone integration was determined from convergence testing to be $\mathrm{10^3}$, $\mathrm{12^3}$, $\mathrm{10^3}$, $\mathrm{10^3}$ and $\mathrm{12^3}$ for Co, Cr, Fe, Mn and Ni, respectively. In the case of matrix-in-impurity calculations, higher $\mathrm{E_{cut}}$ and k-points were chosen. Additionally,3x3x3 supercell containing 108 atoms was employed for FCC-Cr matrix with Co, Fe, Mn and Ni impurity was carried out as well with $\mathrm{\mathrm{E_{cut}}}$ of 700 eV, 600 eV, 600 eV, and 700 eV, respectively. A $\mathrm{10^3}$ k-points grid was used for 3x3x3 supercell calculations. 
\section*{Data and code availability}
Raw data associated with the publication may be provided by authors on request.

\section*{Acknowledgments}
GA is thankful to Phil Hasnip and Keith Refson for helpful discussions, University of Sheffield for PhD scholarship and Leverhulme trust for funding postdoctoral research. GA and CLF are thankful for support provided via our membership
of the UK's HPC Materials Chemistry Consortium, which is funded by EPSRC (EP/L000202). The work of ME has been supported by U.S. Department of Energy, Office of Science, Basic Energy Sciences, Material Sciences and Engineering Division and it used resources of the Oak Ridge Leadership Computing Facility, which is a DOE Office of Science User Facility supported under Contract DE-AC05-00OR22725.
\end{document}


\maketitle

\begin{table}[p]
    \centering
    \begin{tabular}{c|c|c}
        Element &  Pauling scale & Allen's scale \\
        \hline
         Co & 1.88 & 1.84 \\
         Cr & 1.66 & 1.65 \\
         Fe & 1.83 & 1.80 \\
         Mn & 1.55 & 1.75 \\
         Ni & 1.99 & 1.88 
    \end{tabular}
    \caption{Pauling and Allen electrogegativity scale values for Co, Cr, Fe, Mn and Ni taken from \emph{Mann et. al., Journal of the American Chemical Society, 2000, 122, 21,5132-5137}. Note the difference in the electronegativity value of Mn with respect to Cr. Pauling scale predicts Cr to be more electronegative in the comparison with Mn, but Allen's scale predicts the opposite, which is in line with the observation of DFT calculations.}
    \label{tab:electronegativity}
\end{table}

\begin{figure}
    \begin{minipage}{\textwidth}
    \includegraphics[width=\textwidth]{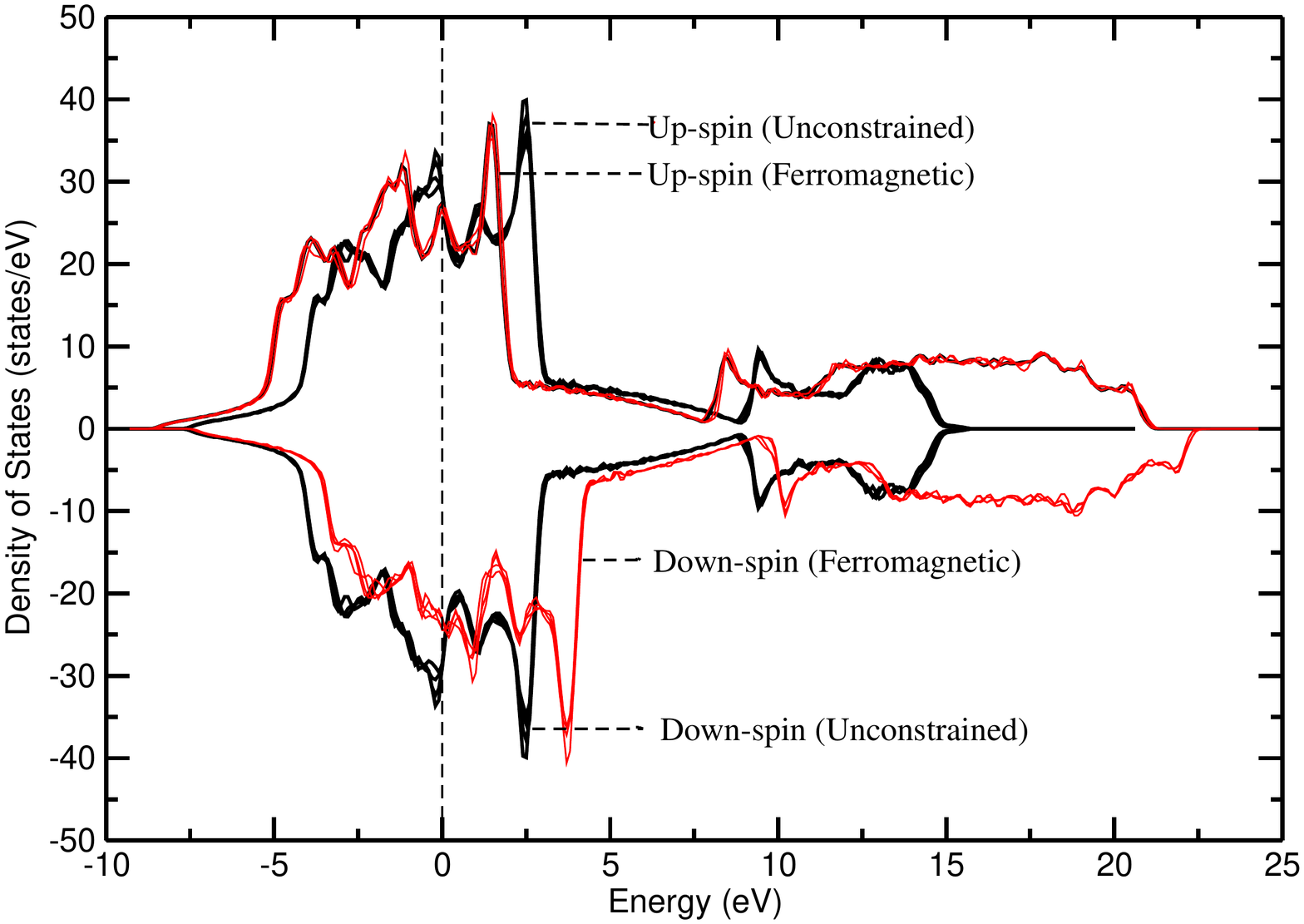}
	    \caption{Total Density of States (DOS) for unconstrained calculation for Co, Fe, Mn and Ni impurity in FCC-Cr matrix, as well as constrained ferromagnetic calculations for each of above-stated cases. Drawn using OptaDOS [\emph{OptaDOS: A Tool for obtaining Density of States, Core-loss and Optical Spectra from Electronic Structure Codes”, Andrew J. Morris, Rebecca J. Nicholls, Chris J. Pickard and Jonathan R. Yates, Comp. Phys. Comm. 185, 5, 1477 (2014)}].} 
    \label{S}
    \end{minipage}
\end{figure}

\begin{figure}
    \centering
    \begin{subfigure}[b]{0.5\textwidth}
        \includegraphics[width=\textwidth]{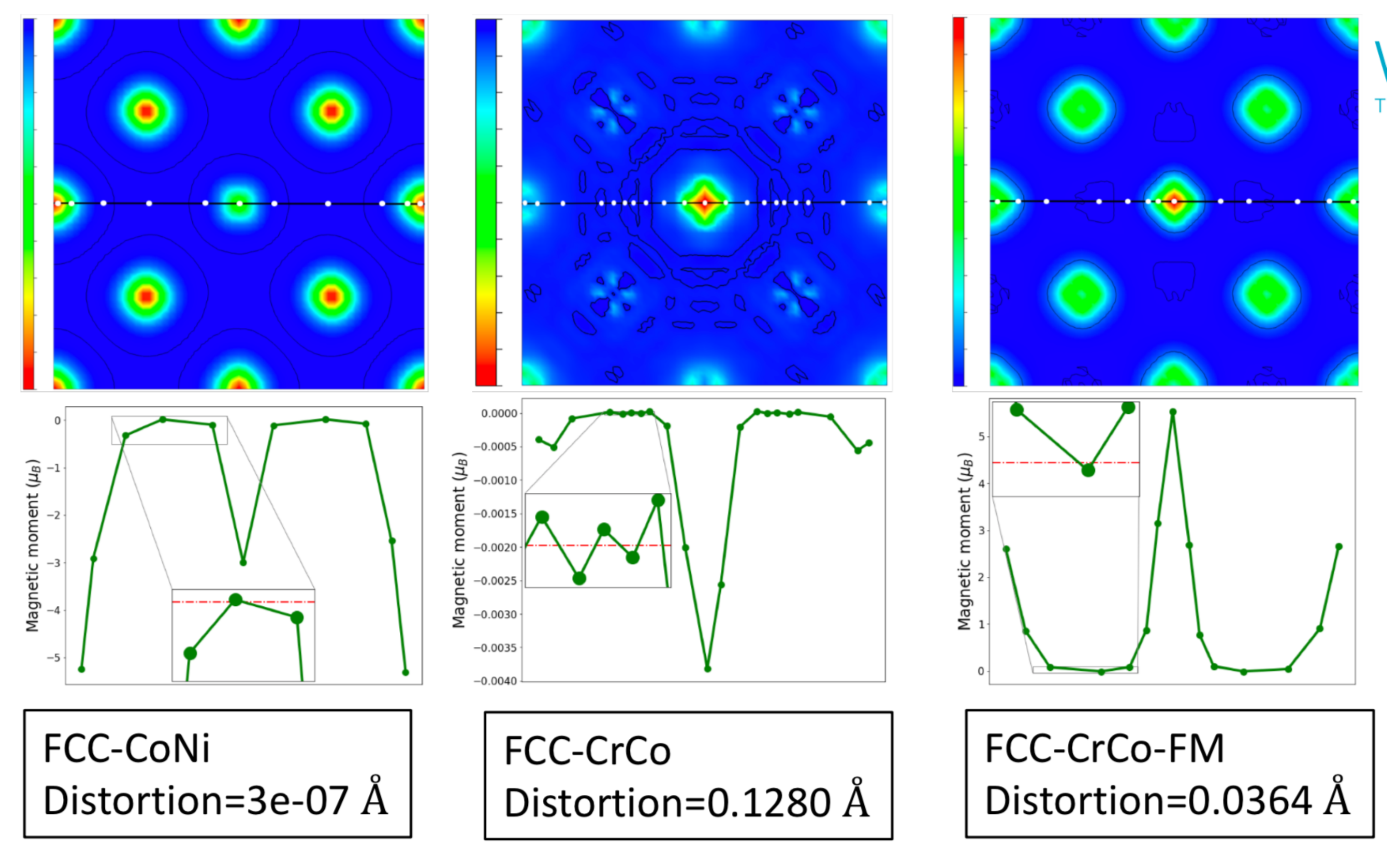}
        \caption{}
        \label{fig1:fluctuation_all}
    \end{subfigure}%
    \begin{subfigure}[b]{0.5\textwidth}
        \stackinset{c}{+0.2in}{t}{-0.4in}{\parbox{3in}{\centering \textbf{FCC-MnCr, high magnetic moment} \newline \textbf{distortion = 0.12 $\mathrm{\AA}$}}}{ 
        \includegraphics[width=\textwidth]{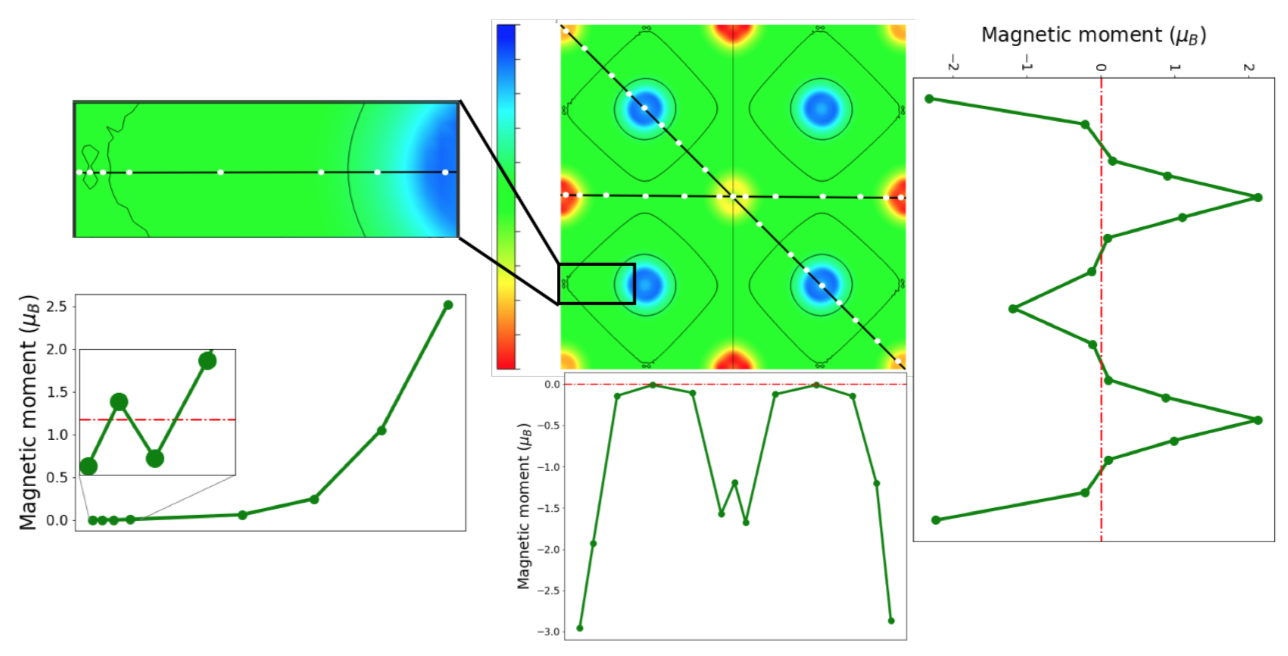}}
        \caption{}
        \label{fig1:Mn-highmm}
    \end{subfigure}
    
    \begin{subfigure}[b]{0.4\textwidth}
        \stackinset{c}{+0.2in}{t}{-0.4in}{\parbox{3in}{\centering \textbf{FCC-MnCr, low magnetic moment} \newline \textbf{distortion = 0.036 $\mathrm{\AA}$}}}{ 
        \includegraphics[width=\textwidth]{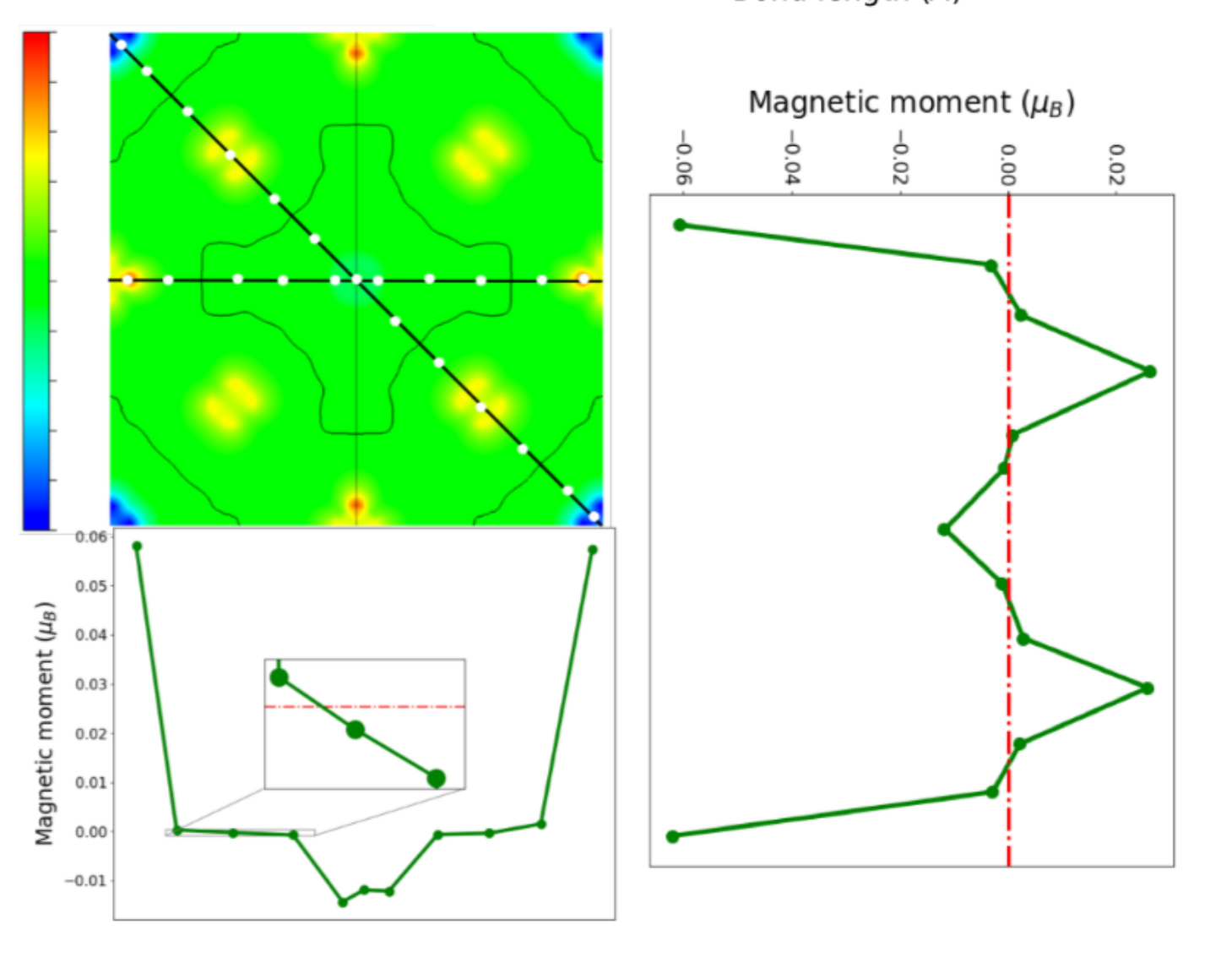}}
        \caption{}
        \label{fig1:Mn-lowmm}
    \end{subfigure}
    
    \caption{
    (\subref{fig1:fluctuation_all}) Variation of magnetic moment in the plane of impurity. The line has been drawn for sake of ease of viewing, where magnetic moment has been taken. It shows the Kohn-Sham orbitals along with the line scan along which we have plotted the variation in the magnetic moment. It can be seen in the case of Ni impurity in FCC-Co, the alignment of spin is in same direction at atomic sites, while in the interstices the direction is opposite, which happens in the case of ferromagnetic systems. For Co impurity in FCC-Cr, the fluctuation in the magnetic moment in the interstitial region. It is interesting to note that when the same system was forced to be ferromagnetic, such fluctuation  in the sign of magnetic moment is absent with corresponding decrease in the distortion in the system. 
    (\subref{fig1:Mn-highmm}) Magnetic moment variation for FCC-MnCr, where high magnetic moment along with high distortion is present.
    (\subref{fig1:Mn-lowmm}) Magnetic moment variation for lower magnetic moment with lower distortion.
    we found two magnetic ground-states for Cr in FCC-Mn. In the first case, lower magnetic moment was seen along with lower distortion (\emph{i.e.,} $\mathrm{0.044\; \AA}$), while in the latter case, higher magnetic moment with higher distortion of $\mathrm{0.124\; \AA}$ was obtained.  We carried out similar line scale in the impurity plane and similar fluctuation in the sigh of magnetic moment can be seen between Mn-Mn bonds, where significant distortion is present (FCC-MnCr, high magnetic moment case)
    }
    \label{S1}
\end{figure}

\begin{figure}
    \centering
    \includegraphics[width=\textwidth]{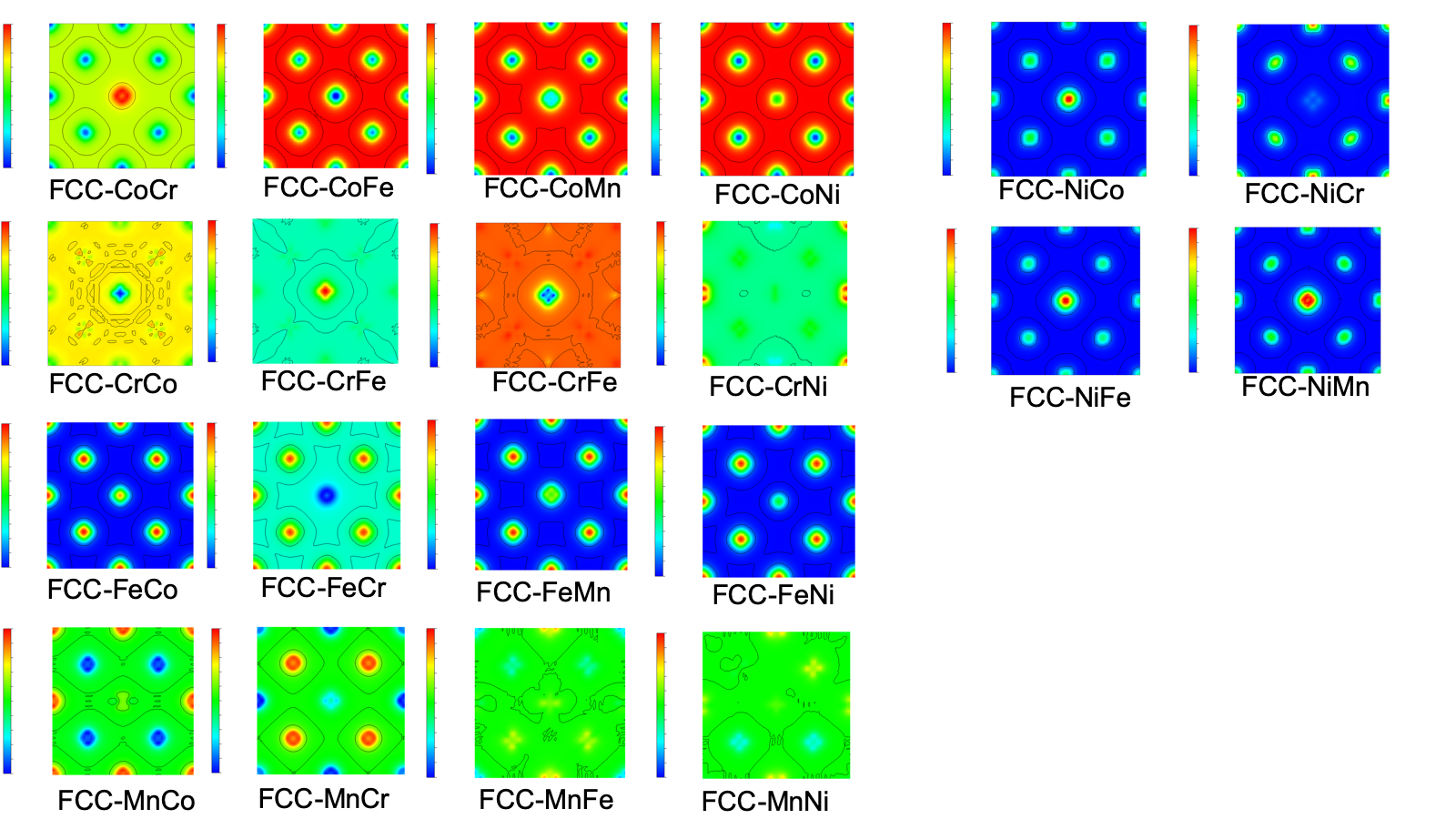}
    \caption{Cross-section of spin-isosurface plot for various cases.}
    \label{S2}
\end{figure}

\begin{figure}
    \centering
    \includegraphics[width=\textwidth]{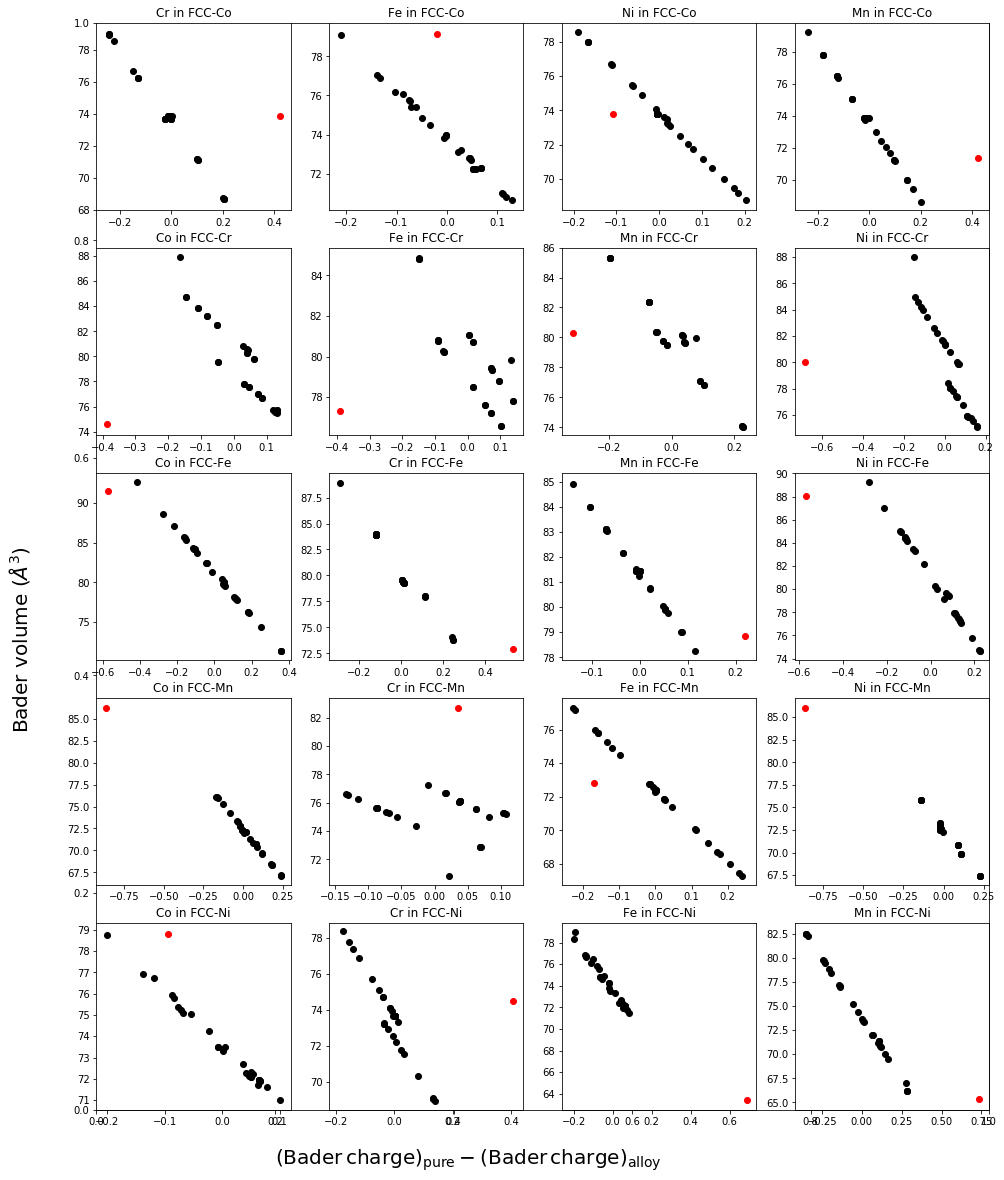}
    \caption{Bader volume and net Bader charge for various cases.}
    \label{S2}
\end{figure}

\begin{figure}
    \centering
    \includegraphics[width=\textwidth]{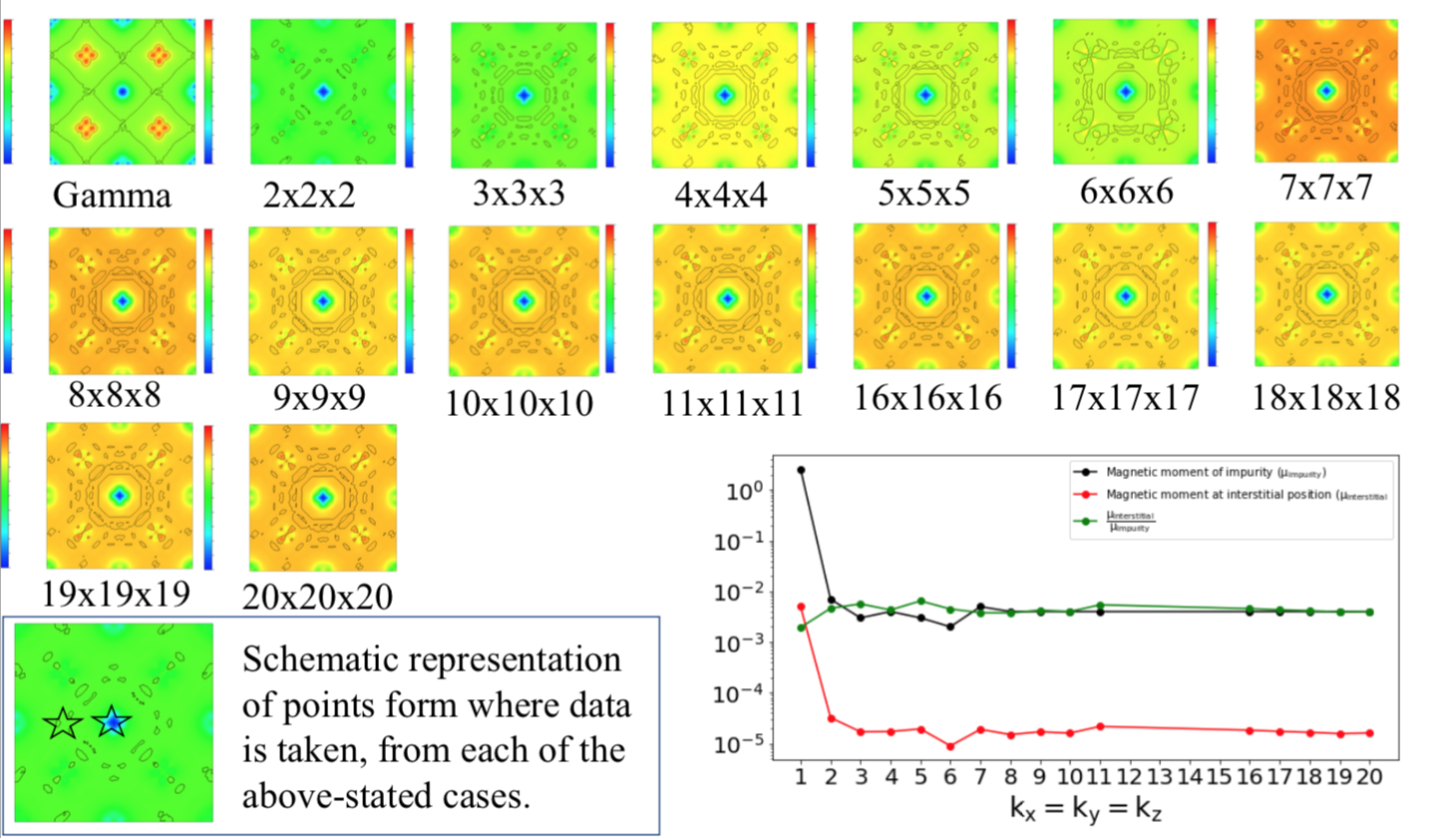}
    \caption{k-point convergence with respect to the magnetic moment of the impurity atom and magnetic moment value at the interstitial position. Note that ratio of magnetic moment at interstitial position to the magnetic moment of the impurity is close to $\mathrm{10^{-3}}$. The 2D cross section for spin density has been drawn using VESTA.}
    \label{S3}
\end{figure}

\begin{figure}
    \centering
    \begin{subfigure}[b]{\textwidth}
        \centering
        \includegraphics[width=\textwidth]{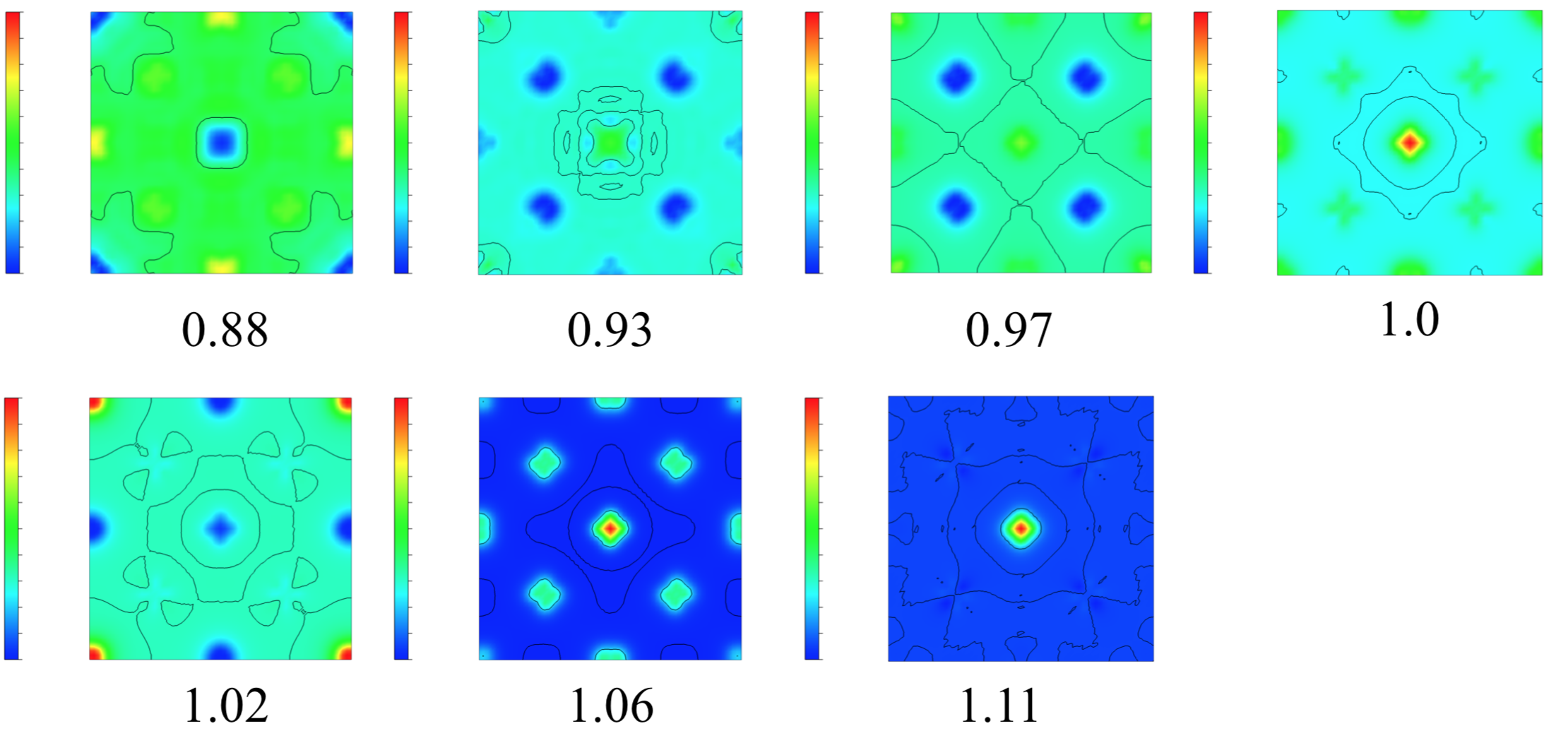}
        \caption{}
        \label{S4-1}
    \end{subfigure}
    \begin{subfigure}[b]{\textwidth}
        \includegraphics[width=\textwidth]{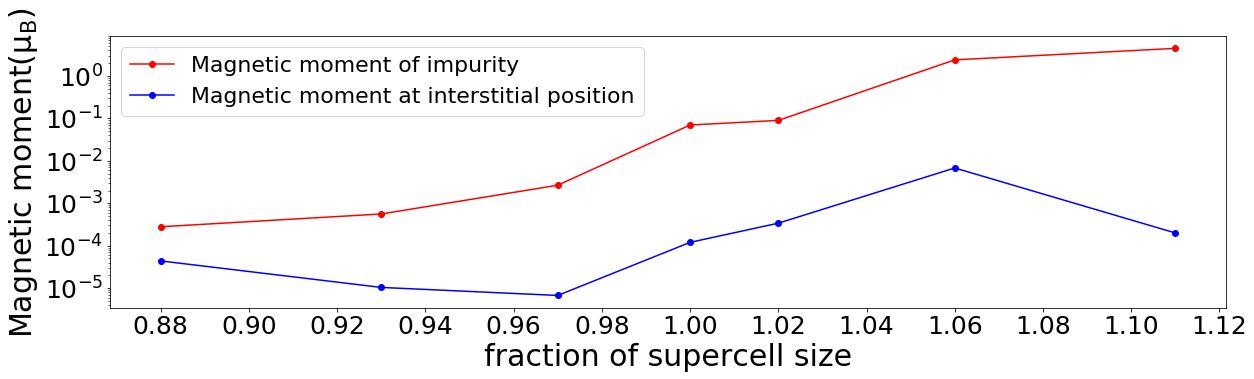}
        \caption{}
        \label{S4-2}
    \end{subfigure}
    \caption{\ref{S4-1}: Mid-section of the supercell of FCC-Cr with Co impurity in the middle. The fraction represent the multiplicative factor with which lattice positions has either been contracted or expanded, \emph{i.e.,} less than one signifies contraction and greater than one implies expansion of the lattice, \ref{S4-2}: Value of magnetic moment for impurity Co atom and magnetic moment at the interstitial site.}
\end{figure}

\begin{figure}
    \centering
    \includegraphics[width=\textwidth]{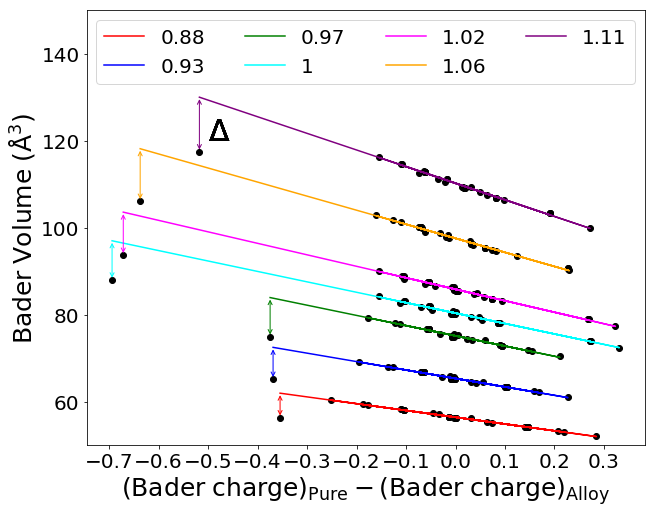}
    \caption{Bader volume variation for impurity atom (Co) and matrix Cr atoms. The $\mathrm{\Delta}$ represents the Bader volume shrink for one of the cases.}
    \label{S5}
\end{figure}

\begin{figure}
    \centering
    \begin{subfigure}[b]{0.45\textwidth}
        \centering
        \includegraphics[width=\textwidth]{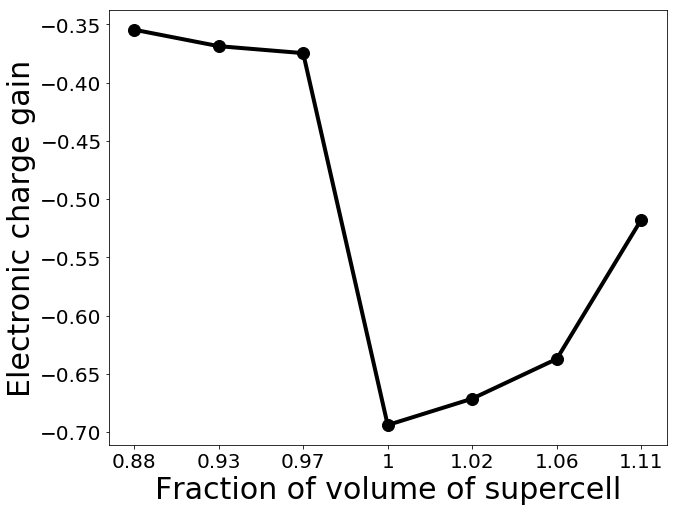}
        \caption{}
        \label{S6-1}
    \end{subfigure}
    \begin{subfigure}[b]{0.45\textwidth}
        \centering
        \includegraphics[width=\textwidth]{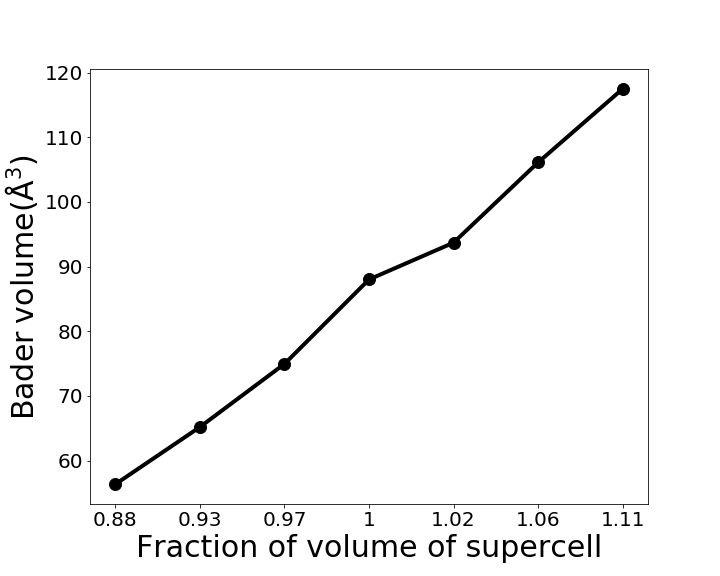}
        \caption{}
        \label{S6-2}
    \end{subfigure}
    \begin{subfigure}[b]{0.45\textwidth}
        \centering
        \includegraphics[width=\textwidth]{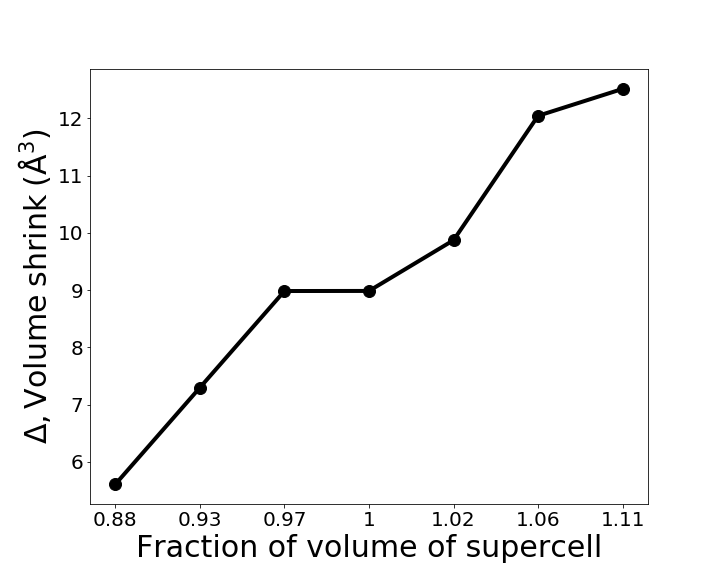}
        \caption{}
        \label{S6-3}
    \end{subfigure}
    \begin{subfigure}[b]{0.45\textwidth}
        \centering
        \includegraphics[width=\textwidth]{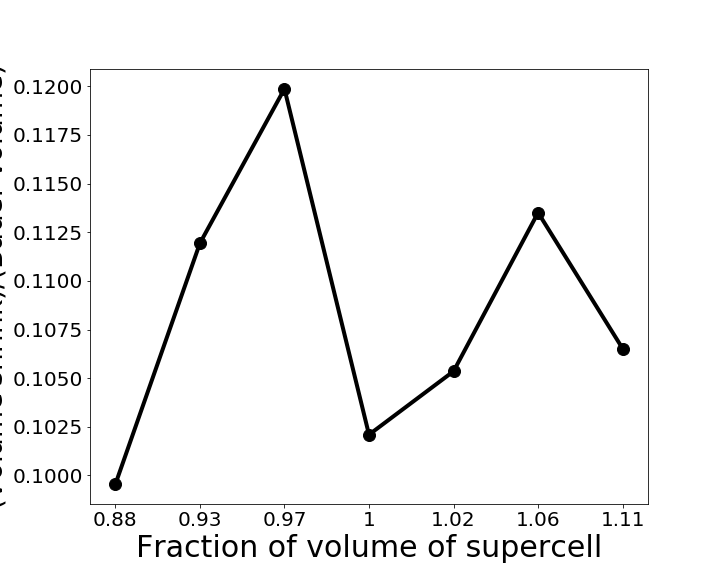}
        \caption{}
        \label{S6-4}
    \end{subfigure}
    \caption{(\ref{S6-1}) Electronic charge gain (or $\mathrm{(Bader\;charge)_{Pure}-(Bader\;charge)_{Alloy}} $) by the impurity Co atom in FCC-Cr matrix, as the supercell is contracted as well as expanded from its equilibrium volume, (\ref{S6-2}) Bader volume of the impurity atom in the centre of the supercell with expansion and contraction of the supercell from equilibrium volume, (\ref{S6-3}) volume shrink of the impurity Co atom with the expansion and contraction of supercell from equilibrium volume and (\ref{S6-4}) ratio of volume shrink and Bader volume with contraction and expansion of the supercell.} 
    \label{S6}
\end{figure}